\documentclass[american,11pt]{article}

\usepackage[T1]{fontenc}
\usepackage[utf8]{inputenc}
\usepackage{lmodern}
\usepackage[margin=1in]{geometry}

\usepackage{amsmath,amssymb,mathtools}
\usepackage{graphicx}
\graphicspath{{FigAS1/}} 
\usepackage{xcolor}
\usepackage{microtype}
\usepackage[american]{babel}
\usepackage{url}
\usepackage{textcomp} 

\usepackage{hyperref}
\hypersetup{
  pdftitle={Quantifying Sand Transport Sensitivity to Dune Shape: Field-Validated CFD with AirSketcher},
  pdflang={en-US},
  pdfkeywords={Coastal dunes, CFD, aeolian transport, dune shape, morphodynamics, wind flow modeling, AirSketcher, sensitivity analysis},
  colorlinks=true, linkcolor=blue, citecolor=blue, urlcolor=blue
}

\title{Quantifying Sand Transport Sensitivity to Dune Shape: Field-Validated CFD with AirSketcher}
\author{%
  Wichai Pattanapol\textsuperscript{1} \and
  Kanisorn Thanutwutthigorn\textsuperscript{2} \and
  Tipaporn Homdee\textsuperscript{3}}
\date{%
  \textsuperscript{1}\textit{Polar Dynamix}\\[-0.25em]
  \textsuperscript{2}\textit{Department of Civil Engineering, Nakhon Phanom University}\\[-0.25em]
  \textsuperscript{3}\textit{Department of Mechanical Engineering, Nakhon Phanom University}\\[0.5em]
  2025-08-28\\
  DOI (this version): \href{https://doi.org/10.5281/zenodo.17157513}{10.5281/zenodo.17157513}\\
  DOI (all versions): \href{https://doi.org/10.5281/zenodo.17095977}{10.5281/zenodo.17095977}
}

\begin{document}
\maketitle

\begin{abstract}
Coastal dune management often alters crest and lee geometry, yet quantifying the transport impact of small shape changes is difficult without heavy models. This paper presents a streamlined, field-anchored CFD workflow (AirSketcher) that ingests a side-profile image, auto-detects the dune outline, and computes near-surface flow. A neutral ABL power-law inlet (\(\alpha=0.16\)) is applied consistently across scenarios; turbulence is closed with a one-equation eddy-viscosity model. Model skill is established against multi-height mast measurements at Tomahawk Beach (Dunedin, NZ): height-matched profile correlations are strong (\(r^2\!\approx\!0.985,\,0.968,\,0.840,\,0.951\)), and the solver reproduces stoss speed-up, crest amplification, and lee-side recovery. For design comparison, a geometry-aware transport proxy is formed as a cubic line-integral of speed along a near-surface polyline (Bagnold-type scaling). Three shapes are evaluated under the validated wind: baseline, top-cut, and back-cut. Integrated proxies (\(\int U^3\,\mathrm{d}s\)) are \(804{,}446\), \(667{,}430\), and \(484{,}779~\mathrm{m}^4\,\mathrm{s}^{-3}\), implying \(\approx\!-17\%\) (top-cut) and \(\approx\!-40\%\) (back-cut) vs.\ baseline. Flow patterns explain the reductions—crest-peak attenuation, muted lee jets, earlier reattachment—concentrating deposition nearer the crest, especially for the back-cut. The approach offers a rapid, interpretable metric for screening dune modifications before committing to fully coupled morphodynamic modeling.

\end{abstract}

\section{Introduction}\label{introduction}
Aeolian sand transport scales strongly with near-surface wind speed. Classical arguments indicate that the cubic dependence on flow speed amplifies high-velocity pockets over crests and in lee-side acceleration zones, making geometry a primary control. This study couples field measurements with a streamlined Computational Fluid Dynamics (CFD) workflow (AirSketcher) to (i) validate flow predictions and (ii) compare three candidate dune modifications using a transport proxy derived from a Bagnold-type law.

\section{Methods}\label{methods}

\subsection{Field measurements (overview)}\label{field-measurements-overview}
A focused field campaign was conducted at Tomahawk Beach (Dunedin City, New Zealand) along a foredune transect directly implicated in proposed sand-removal works [15]. The foredune face is steep (\(\approx 70^\circ\)), with sparse stoss vegetation and denser cover beyond the crest (marram grass, New Zealand flax, and scattered shrubs/trees). This configuration accentuates near-surface acceleration on the stoss slope, crest speed-up, and lee-side recirculation—features commonly documented in hill/dune aerodynamics and desirable for validating a CFD representation of the flow [2, 10].

\begin{figure}
\centering
\includegraphics[width=0.85\linewidth,keepaspectratio]{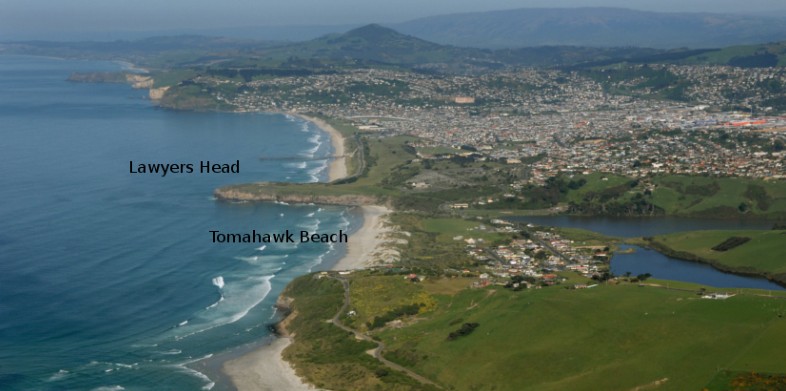}
\caption{Tomahawk case study: Lawyers Head. Beyond the backdune at Tomahawk is a football field and the residential suburb of Tomahawk.}
\end{figure}

Regional wind climatology from the Lawyers Head mast shows frequent but weaker northerlies and less frequent, stronger south-westerlies; sand-transport potential is therefore dominated by southerly-sector events. The campaign was timed in July 2008 during a southerly episode that oriented the flow approximately perpendicular to the transect, enhancing the diagnostic signal in the measurements. A practical reference velocity of about 22~m\,s\(^{-1}\) at \(z=5\)~m above the beach was documented during this period, with higher gusts recorded [15].

Topography was surveyed with a Leica total station along a single cross-shore line spanning the upper beach, foredune toe, stoss, crest, lee, and backdune. The selected reach is topographically simple (few blowouts) and coincides with the management focus area. The surveyed ground profile provides the geometric backbone for both meshing and the placement of CFD line probes, and ensures that measured anemometer heights are referenced to the local surface along the transect.

\begin{figure}
\centering
\includegraphics[width=0.85\linewidth,keepaspectratio]{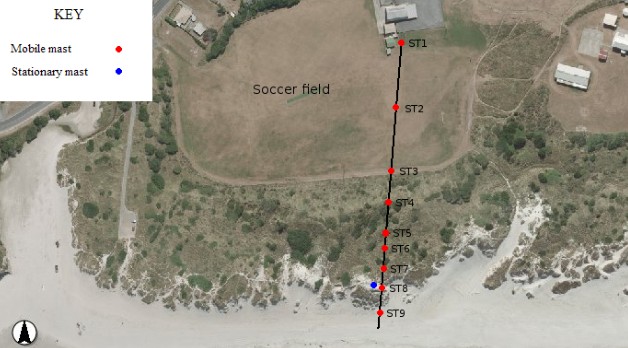}
\caption{Transect line with locations of mobile (ST 1--7 and 9), and stationary masts (ST 8) at Tomahawk Beach.}
\end{figure}

Nine stations (ST1--ST9) were deployed to sample key aerodynamic zones—beach approach, foredune toe, upper stoss/crest, and multiple lee and backdune positions. A mobile mast at each station carried five cup anemometers at nominal heights of 0.2, 0.5, 1.0, 2.0, and 5.0~m above ground, plus a wind vane at 2.0~m for direction [15]. The mast dwelled for five minutes per station, logging every three seconds with a Campbell 21X Micrologger. To track background variability during redeployments, a stationary reference mast near the stoss crest carried an R.M. Young anemometer at 3.0~m, providing a concurrent record of speed and direction.

Vegetation structure along the transect was mapped into three canopy zones commonly used at this site: \(\approx\)0.5~m marram on the back of the foredune, \(\approx\)1.0~m in the lee depression, and \(\approx\)0.5~m on the backdune. These observations support roughness specification and, where required, canopy-drag parameterization in the numerical model, consistent with boundary-layer micrometeorology practice [4, 12].

Sampling followed a sequential station order to capture spatial gradients under quasi-steady synoptic forcing. Standard quality control included pre/post-campaign checks of mounting height and alignment, trimming of handling transients, conservative despiking, and computation of 5–10 minute means (with standard deviations) per height. All heights were referenced to surveyed ground elevations so that comparisons with CFD probes at fixed offsets above the terrain are direct and unambiguous.

For the CFD boundary condition, a southerly inlet consistent with field conditions was prescribed using the reference velocity (about 22~m\,s\(^{-1}\) at \(z=5\)~m) and a neutral ABL profile following standard log-law formulations [4, 12]. Model–data comparisons emphasize stoss speed-up, crest amplification, and lee-side recovery extracted from AirSketcher line probes aligned with the surveyed transect, in line with established hill/dune-flow behavior [2, 10].

\subsection{CFD tool (AirSketcher)}\label{cfd-tool-airsketcher}
AirSketcher is used here as a two-dimensional, incompressible RANS solver with an eddy-viscosity closure and a projection step for pressure--velocity coupling (see general turbulence/near-wall background in [11]). Only the components described below are employed in this study.

\textit{Governing equations.} For constant density \(\rho\), the model advances velocity \(\vec{U}=(u,v)\) and pressure \(p\):
\[
\nabla\cdot\vec{U}=0,\qquad
\frac{\partial \vec{U}}{\partial t}
+(\vec{U}\cdot\nabla)\vec{U}
= -\frac{1}{\rho}\nabla p
+ \nabla\cdot\!\Big[(\nu+\nu_t)\big(\nabla\vec{U}+(\nabla\vec{U})^{T}\big)\Big],
\]
with molecular viscosity \(\nu\) and eddy viscosity \(\nu_t\).

\textit{Discretisation.} Advection is advanced explicitly with a variable time step \(\Delta t\); diffusion and model terms use second-order central differences. A bounded upwind/central blend is applied to convective terms to control spurious oscillations while preserving attached-flow gradients.

\textit{Pressure--velocity coupling (projection).} A provisional velocity \(\vec{U}^{\*}\) is formed from the advection--diffusion step. The pressure correction \(\phi\) solves
\[
\nabla^{2}\phi=\frac{\rho}{\Delta t}\,\nabla\cdot\vec{U}^{\*},
\]
followed by
\[
\vec{U}^{\,n+1}=\vec{U}^{\*}-\frac{\Delta t}{\rho}\nabla \phi,\qquad
p^{\,n+1}=p^{\,n}+\phi.
\]

\textit{Turbulence closure (Spalart--Allmaras).} Turbulence is represented with the one-equation SA model for the working variable \(\tilde{\nu}\) [3]:
\[
\frac{\partial \tilde{\nu}}{\partial t}
+\vec{U}\cdot\nabla \tilde{\nu}
= C_{b1}(1-f_{t2})\,\tilde{S}\,\tilde{\nu}
+ \frac{1}{\sigma}\!\left[\nabla\cdot\big((\nu+\tilde{\nu})\,\nabla \tilde{\nu}\big)
+ C_{b2}\,|\nabla \tilde{\nu}|^{2}\right]
- C_{w1}\,f_w\!\left(\frac{\tilde{\nu}}{d}\right)^{2},
\]
with eddy viscosity
\[
\nu_t=\tilde{\nu}\,f_{v1},\qquad
f_{v1}=\frac{\chi^{3}}{\chi^{3}+C_{v1}^{3}},\quad
\chi=\frac{\tilde{\nu}}{\nu}.
\]
Here \(d\) is the wall distance; \(f_{t2}\), \(f_w\) and \(\tilde{S}\) are the standard SA functions. Canonical coefficients are used; near-ground shear is resolved on the given mesh (no wall functions).

\textit{Domain and boundaries.} The computational plane follows the surveyed transect; the dune surface is imposed from the total-station profile on a uniform Cartesian grid.
\begin{itemize}
\item Inlet (neutral ABL):
\(U(z)=U_{\mathrm{ref}}\,\dfrac{\ln((z+z_0)/z_0)}{\ln((z_{\mathrm{ref}}+z_0)/z_0)}\),
with \(U_{\mathrm{ref}}\) specified at \(z_{\mathrm{ref}}=5~\mathrm{m}\) and roughness length \(z_0\) consistent with neutral log-law practice [12, 4].
\item Outlet: zero gradient for \(\vec{U}\) and the pressure correction.
\item Top: symmetry/free-slip (\(U_n=0,\ \partial U_t/\partial n=0\)).
\item Ground and dune surface: no-slip (\(\vec{U}=\vec{0}\)).
\end{itemize}

\textit{Time-step control (CFL).}
\[
\mathrm{CFL}=\max_{i,j}\!\left(\frac{|u|\,\Delta t}{\Delta x}+\frac{|v|\,\Delta t}{\Delta y}\right)\le \mathrm{CFL}_{\max},\qquad
\Delta t=\eta\,
\min\!\left(\frac{\Delta x}{|u|+\epsilon},\,\frac{\Delta y}{|v|+\epsilon}\right),
\]
with small \(\epsilon\) for robustness and \(\eta\) derived from \(\mathrm{CFL}_{\max}\).

\textit{Mass balance monitoring.}
\[
\dot{m}=\int_{\Gamma}\rho\,\vec{U}\cdot\vec{n}\,\mathrm{d}A,\qquad
\delta_m=\frac{|\dot{m}_{\mathrm{in}}-\dot{m}_{\mathrm{out}}|}{\dot{m}_{\mathrm{in}}}\times 100~\%.
\]

\textit{Polyline post-processing (transport proxy).}
Near-surface speed \(U\) is sampled along user-defined polylines coincident with the transect. A Bagnold-type cubic proxy is formed for comparative transport potential,
\[
Q_{\mathrm{line}}=C\int_{s_0}^{s_1} U(s)^{3}\,\mathrm{d}s,
\]
with dimensionless \(C\) empirical. The cubic weighting follows classical aeolian arguments [1] and modern summaries of wind-blown sand physics [9]. For context, \(\int U\,\mathrm{d}s\) and \(\int U^{2}\,\mathrm{d}s\) are also reported, together with \(p(s)\) and the components \(u(s),v(s)\), to relate speed-up, pressure deficit, and directionality to dune shape.

\subsection{Domain and dune geometries}\label{domain-and-dune-geometries}
The measured dune profile was digitized from the total-station survey and mapped into model coordinates; the beach datum defines \(y=0\) at the shoreline foot of the foredune [15]. Within this box, three geometries were analyzed:
\begin{itemize}
\item S1 — Baseline (BL): unmodified survey profile.
\item S2 — Top-cut: crest lowering over a finite crest window, followed by short-scale smoothing on both stoss and lee to avoid artificial corners.
\item S3 — Back-cut: lee-side trimming (crest shifted landward) with gentle re-grading of the lee slope to the original backdune tie-in.
\end{itemize}
All modified shapes preserve beach and backdune tie-points so that inlet/outlet sections remain identical across cases; differences in flow and transport are therefore attributable to geometry rather than domain placement.

\subsection{Transport proxy from line probes (Bagnold-type)}\label{transport-proxy}
A near-surface velocity profile is obtained by drawing a polyline line probe that follows the ground (or a specified offset height). The solver samples the local speed \(U(s)=\sqrt{u(s)^2+v(s)^2}\) along the polyline, parameterized by arc length \(s\), and computes segment lengths \(\Delta s_i\) between successive sample points. From this measured profile, a cubic line-integral is formed and used as a proxy for potential sand transport in the Bagnold sense [1].

\textit{Raw cubic line-integral (from the measured line probe).}
\[
Q_{\mathrm{line}}
=\int_{s_0}^{s_1} U(s)^3\,\mathrm{d}s
\;\approx\;
\sum_{i=1}^{N-1}\frac{U_i^3+U_{i+1}^3}{2}\,\Delta s_i.
\]
This directly reflects what the probe measured (no threshold), so it is useful for like-for-like comparison between shapes on the same transect.

\textit{Bagnold-type, thresholded cubic form (used as transport estimate).}
\[
Q_{\mathrm{Bag}}
= C \int_{s_0}^{s_1} \big[\max(U(s)-U_t,\,0)\big]^3\,\mathrm{d}s
\;\approx\;
C \sum_{i=1}^{N-1}
\frac{\big[\max(U_i-U_t,0)\big]^3+\big[\max(U_{i+1}-U_t,0)\big]^3}{2}\,\Delta s_i.
\]
Here \(U_t\) is the site-specific threshold speed (linked to threshold friction velocity \(u_{*t}\)) and \(C\) is an empirical, dimensionless coefficient; common ranges for \(U_t\)/\(u_{*t}\) and their dependence on grain size and density are discussed in [7, 8, 9]. There is no universal \(C\); a single calibrated value is held fixed across scenarios for fair comparison.

\textit{Rationale.}
The cubic dependence emphasizes high-speed patches over crests and in jets, consistent with Bagnold-type sensitivity to near-surface momentum [1] and with modern saltation physics syntheses [9, 11]. Using the measured polyline profile ensures the integral reflects the actual flow response to geometry; co-variations with along-probe pressure \(p(s)\) and components \(u(s),v(s)\) help relate speed-up, lee-side deficits, and secondary flows to potential transport pathways (cf.\ observations around dune flanks and wakes in [10]).

\section{Results}\label{results}

\subsection{Validation against field data}\label{validation}
\begin{figure}
\centering
\includegraphics[width=0.9\linewidth,keepaspectratio]{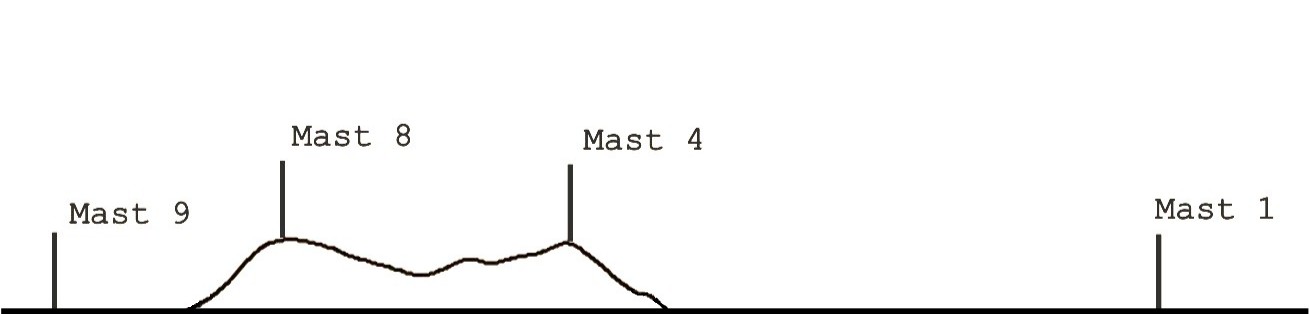}
\caption{Mast locations (M1, M4, M8, M9).}\label{fig:mastposition}
\end{figure}

\begin{figure}
\centering
\includegraphics[width=0.9\linewidth,keepaspectratio]{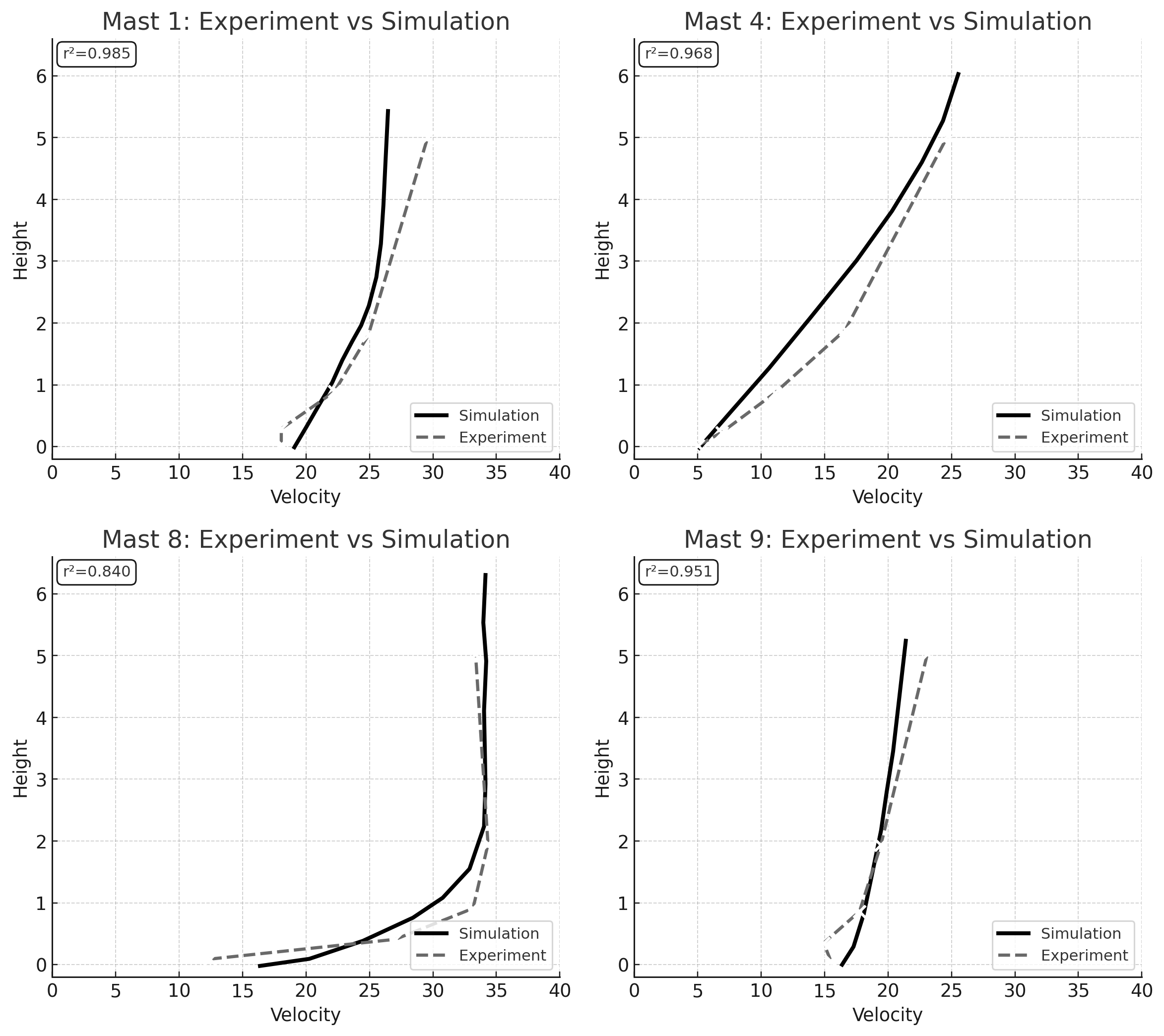}
\caption{Vertical wind-speed profiles at Masts 1, 4, 8, 9.}\label{fig:mastmatrix}
\end{figure}

Spatial context (Fig.~\ref{fig:mastposition}) shows the relative mast locations within the flow field and the local speed-up/slow-down regions sampled by each mast.

Vertical-profile comparisons (Fig.~\ref{fig:mastmatrix}): Agreement is strong at masts 1, 4, and 9, and moderate at mast 8. Height-matched \(r^2\) (simulation interpolated to instrument heights):
\[
\text{M1: }0.985;\quad \text{M4: }0.968;\quad \text{M8: }0.840;\quad \text{M9: }0.951.
\]
Interpretation: The model reproduces vertical shear and local speed-ups well at masts 1, 4, and 9. Mast 8 shows comparatively lower agreement, suggesting a site-specific mismatch (e.g., inlet specification, local roughness/porosity, or unresolved blockage) that warrants targeted calibration and/or additional measurements.

\subsection{Flow over modified shapes (S1--S3)}\label{flow}
\begin{figure}
\centering
\includegraphics[width=\linewidth,keepaspectratio]{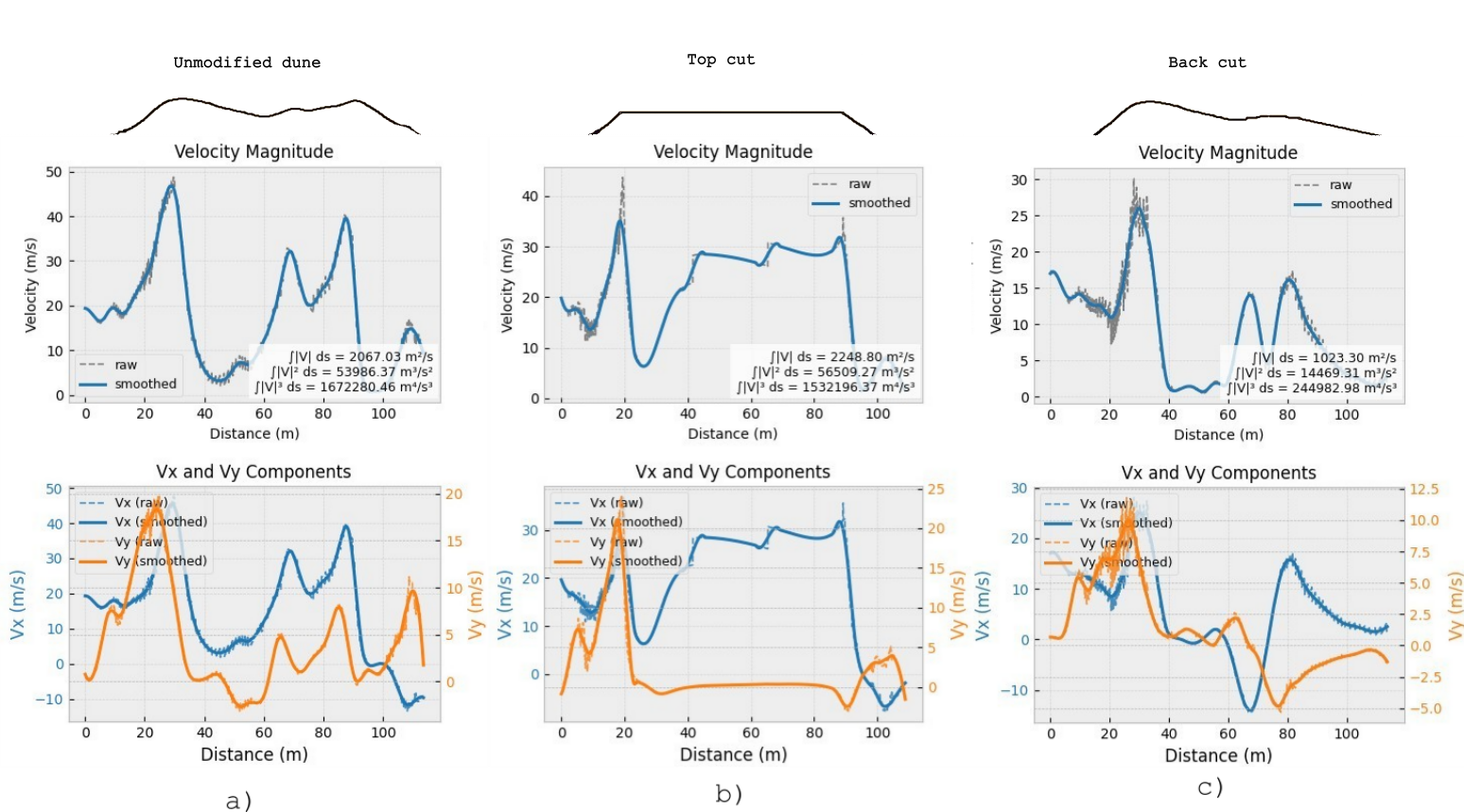}
\caption{Velocity profiles along the near-surface transect: (a) original dune, (b) top-cut, (c) back-cut.}\label{fig:vxvy}
\end{figure}

Figures~\ref{fig:vxvy}a--c correspond to scenario S1 (original terrain), S2 (top-cut modification), and S3 (back-cut modification). In all cases, the plotted variable is the velocity magnitude \(|U|\) (flow speed), sampled along a polyline positioned a small distance above the surface to avoid the solid boundary.

\begin{figure}
\centering
\includegraphics[width=0.9\linewidth,keepaspectratio]{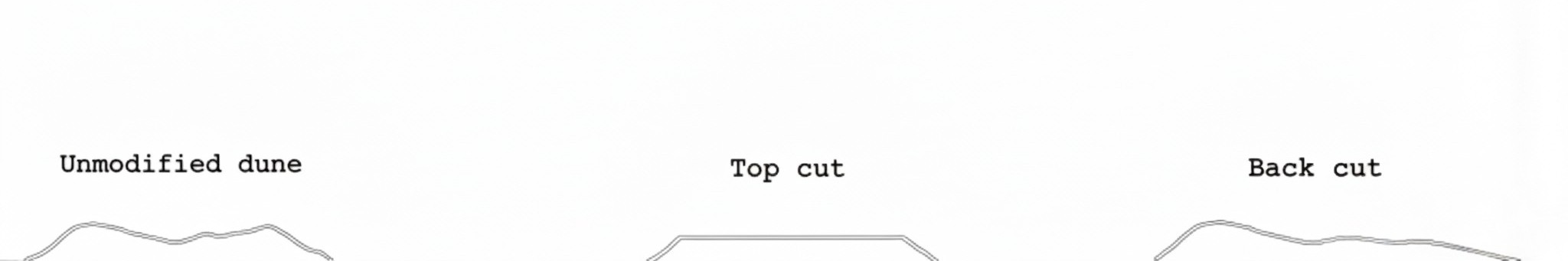}
\caption{Dune modification scenarios: S1 unmodified dune, S2 top cut and S3 back cut.}\label{fig:streamline}
\end{figure}

\medskip
Key observations for the validated wind:
\begin{itemize}
\item S1 (original; Fig.~\ref{fig:vxvy}a): strong crest speed-up (largest \(|U|\) peak); sharp pressure drop at the crest; pronounced lee jet and later reattachment (negative \(v_x\)), with larger fluctuations in the segment-wise mass flux.
\item S2 (top-cut; Fig.~\ref{fig:vxvy}b): crest flattening reduces peak acceleration and smooths the pressure trough; the lee jet is weaker and separation length shortens; fluctuations in mass flux are damped relative to S1.
\item S3 (back-cut; Fig.~\ref{fig:vxvy}c): lee truncation shifts the crest downwind; stoss-side acceleration occurs earlier but the lee jet is most muted; earliest reattachment among the three, with the smallest post-crest \(|U|\) peak.
\end{itemize}

\subsection{Transport comparison via \texorpdfstring{\(\int U^{3}\,\mathrm{d}s\)}{\textbackslash int U\^{}3 ds}}\label{transport-comparison}
Using the validated setup and a consistent near-surface polyline just above the ground, bulk sand transport under super-threshold wind is taken to scale with the cubic speed integral, \(\propto \int U^{3}\,\mathrm{d}s\), following Bagnold-type arguments and later syntheses of saltation physics [1, 9, 11].

\begin{itemize}
\item Integrated proxies (absolute, \(\mathrm{m}^4\,\mathrm{s}^{-3}\)): S1 = \textbf{804\,446}, S2 = \textbf{667\,430}, S3 = \textbf{484\,779}.
\item Normalized to baseline (S1 = 1.00): \(Q_{\text{proxy}}/Q_{\text{S1}}=\{\,\textbf{1.000},\,\textbf{0.830},\,\textbf{0.603}\,\}\).
\item \(\Delta\) vs S1: S2: \textbf{--17.0\%}, S3: \textbf{--39.7\%}.
\item Ranking: S3 $<$ S2 $<$ S1.
\end{itemize}

Notes on interpretation and robustness: normalization removes dependence on the empirical prefactor \(C\) and reduces sensitivity to threshold choices \(U_t\) (or \(u_{*t}\)) used in thresholded variants, which mainly shift absolute magnitudes [1, 7, 8, 11]. The observed ordering aligns with classic hill-flow expectations—crest-peak attenuation, weaker lee jets, and earlier reattachment in S2/S3 reduce high-velocity patches that disproportionately weight the cubic integral [2].

\subsection{Sensitivity and uncertainty}\label{sensitivity}
This study uses a thresholded cubic proxy
\[
Q \propto \int [\max(U-U_t,0)]^3\,\mathrm{d}s,
\]
with \((x)_+=\max(x,0)\) along the near-surface sampling polyline.

\begin{itemize}
\item Threshold \(U_t\). Applying a speed threshold—either directly via \(U_t\) or via a threshold friction velocity \(u_{*t}\)—down-weights marginal/recirculation zones and lowers absolute totals:
\[
Q(U_t)\propto\int [\max(U-U_t,0)]^3\,\mathrm{d}s,\qquad
Q(u_{*t})\propto\int [\max(u_*-u_{*t},0)]^3\,\mathrm{d}s,
\]
with neutral log-law \(u_*=\kappa\,U(z)/\ln(z/z_0)\). Across \(U_t\in[4,7]\) m\,s\(^{-1}\) the ordering is unchanged (\(S3<S2<S1\)); normalized ratios decrease slightly as \(U_t\) increases, consistent with threshold formulations [7--9].
\item Empirical coefficient \(C\). Acts as a global scalar (\(Q\propto C\)). It cancels in relative comparisons (e.g., \(Q_i/Q_j\)), affecting only absolute magnitudes (cubic Bagnold-type scaling [1, 9, 11]).
\item Probe/trace height. Offsetting the sampling height modifies \(U(z)\) via neutral-ABL mapping (log-law or power-law), so absolute \(Q\) changes monotonically but the ranking remains stable [4--6, 12, 13].
\item Surface roughness \(z_0\). If a common \(z_0\) is used for all scenarios, its effect largely cancels in ratios. Shape-specific \(z_0\) rescales \(U(z)\) (and thus \(u_*\)) through the velocity–roughness relation, changing absolute \(Q\).
\item Air density \(\rho_a\). Linear scaling: \(Q\propto\rho_a\). Relative comparisons are insensitive when \(\rho_a\) is shared across runs (aeolian transport scalings [1, 9]).
\item Smoothing / discretization. Smoothed and raw polylines yield consistent integrals; smoothing removes high-frequency noise without materially changing \(\int(\cdot)^3\,\mathrm{d}s\). With trapezoidal quadrature on segments \(\Delta s\),
\[
Q \approx C\,\rho_a\,\sum_k \frac{\Delta s_k}{2}\Big(\,[\max(U_k-U_t,0)]^3+[\max(U_{k+1}-U_t,0)]^3\Big),
\]
introducing only small numerical uncertainty that does not alter conclusions.
\end{itemize}

Normalization: For case \(i\), define \(Q_i^{\mathrm{rel}}=Q_i/Q_{\mathrm{ref}}\). Global factors shared across cases (e.g., \(C\), \(\rho_a\), common \(z_0\)) cancel in \(Q_i^{\mathrm{rel}}\), and parameters that mainly reshape \(U(z)\) preserve the observed ordering.

\section{Discussion}\label{discussion}
The cubic polyline integral offers a fast, geometry-aware discriminator of aeolian transport potential that preserves the Bagnold-type cubic sensitivity to high-speed pockets over crests and in lee jets [1, 9, 11]. In practice it remains robust under plausible choices of threshold \(U_t\), probe height, and roughness \(z_0\), reflecting well-understood neutral-ABL profile behavior and threshold physics [4--7, 12]. Applied to the Tomahawk transect, both modifications reduce the proxy relative to the original dune: the top-cut lowers it by about 17\%, while the back-cut lowers it by about 40\%. These reductions align with canonical flow features—crest-peak attenuation, weaker lee jets, and earlier reattachment—that are known to diminish near-surface speeds downstream of the crest [2, 10, 14]. Absolute transport estimates require a site-specific coefficient \(C\) and a threshold/friction-velocity treatment (e.g., \(u_{*t}\)), but the relative ranking is insensitive to those choices under super-threshold winds [1, 7, 9]. Consequently, the integral provides a reliable rapid-iteration metric for shape comparison prior to higher-fidelity morphodynamic modeling or fully coupled saltation simulations [11].

\section{Conclusions}\label{conclusions}
\begin{enumerate}
\item Validated flow patterns. AirSketcher reproduces the key dune-flow signatures and their geometric controls. In the scenario runs, S2 (top-cut) lowers the crest peak and shortens separation, while S3 (back-cut) shows the earliest reattachment (negative velocity in Fig.~\ref{fig:vxvy}c) and the weakest lee jet. Both modifications attenuate the high-velocity pockets that dominate cubic scaling.
\item Transport proxy is geometry-sensitive and robust. A Bagnold-type cubic line-integral along a near-surface polyline provides a stable, comparable proxy across shapes; rankings are insensitive to reasonable choices of threshold, probe height, roughness, and smoothing. Using the measured polylines, the integrated proxies yield S3 $<$ S2 $<$ S1 (about 40\% and 17\% below the original, respectively).
\item Where will sand settle (flow-pattern view). Deposition is favored where speed decays rapidly downwind of the crest, i.e., where the along-probe flux gradient is negative: \(-\partial q/\partial x>0\) with \(q \propto U^{3}\).\\
S3 (back-cut): the muted jet and earliest reattachment create the steepest post-crest decay in \(U\), concentrating deposition just lee of the crest and within the cut; this promotes local ``self-healing.''\\
S2 (top-cut): peak attenuation and a shortened wake also enhance near-crest decay, but the lee jet is not as suppressed as in S3. Deposition shifts upwind relative to S1, yet some sand is carried farther into the lee compared with S3.
\item How fast will S2/S3 fill back in (qualitative expectation). Actual rates depend on supply (dry fetch, tide/moisture), event frequency/intensity, and vegetation. Holding wind forcing equal, the expected local infill tendency ranks
\[
\text{infill speed (local)}:\quad S3 \gtrsim S2 \gg S1.
\]
S3 should accumulate fastest near the cut (earliest reattachment, strongest near-crest sink), often over fewer energetic events. S2 should accumulate at a moderate pace, with deposition centered near/just lee of the crest but with a portion advected farther downwind than in S3. S1 shows the slowest near-crest accumulation because the stronger lee jet transports sand deeper into the lee before it settles.
\item Design/maintenance implication. If the goal is to minimize export along the transect, S3 offers the largest reduction in the transport proxy and the strongest near-crest trapping. S2 still reduces export versus the original and should infill, but more gradually and with some deposition displaced farther downwind. For absolute timelines, pair these flow-based expectations with site supply metrics (dry-beach fetch frequency, wind roses, moisture/threshold exceedance, and vegetation establishment).
\end{enumerate}

\section{Practical recipe (reproducibility)}
AirSketcher uses a side-on image (no DEM required). The software detects the dune outline from an imported profile image and runs the flow. Once a polyline probe is drawn, the software automatically computes all along-line integrals—including \(\int U^3\,\mathrm{d}s\)—so no external post-processing is needed.
\begin{enumerate}
\item Import profile image \(\rightarrow\) detect contour. Load a side-profile photo/drawing (PNG/JPG). Use Detect Contour / edge tools to auto-trace the dune; tidy with edit handles. Set the length scale with a known distance (e.g., staff marks). Save as Baseline; duplicate and edit to create Top-cut and Back-cut (keep beach/backdune tie-points aligned).
\item Set inlet profile (kept identical across scenarios). Use the ABL (power-law) inlet with \(\alpha=0.16\) and a specified reference speed/height pair. (When the log-wind profile is selected, AirSketcher assigns surface roughness automatically; in these runs, the power-law ABL with \(\alpha\) represents the vertical shear.) Hold inlet type and parameters identical for Baseline, Top-cut, and Back-cut.
\item Run to steady state. Monitor residuals and mass-balance with the same convergence criteria across cases.
\item Automatic integrals. Along the probe, AirSketcher reports \(\int U\,\mathrm{d}s\), \(\int U^2\,\mathrm{d}s\), \(\int U^3\,\mathrm{d}s\) (the Bagnold-style proxy), plus length-normalized badges and per-segment diagnostics. Export CSV/plots if needed.
\item Compare and rank. Use the Normalized view to get relative factors for Top-cut and Back-cut. In this study: \(-17\%\) and \(-40\%\) vs.\ Baseline, respectively.
\end{enumerate}

\section*{References}\label{bibliography}
\begin{enumerate}
\setlength{\itemsep}{0.25\baselineskip}%
\renewcommand{\labelenumi}{[\theenumi]}%
  \item R. A. Bagnold, \emph{The Physics of Blown Sand and Desert Dunes}. Methuen, 1941.
  \item P. S. Jackson and J. C. R. Hunt, “Turbulent wind flow over a low hill,” \emph{QJRMS}, 101(430), 929--955, 1975.
  \item P. R. Spalart and S. R. Allmaras, “A one-equation turbulence model for aerodynamic flows,” in \emph{30th AIAA Aerospace Sciences Meeting}, 1992.
  \item R. B. Stull, \emph{An Introduction to Boundary Layer Meteorology}. Kluwer, 1988.
  \item C. G. Justus, W. R. Hargraves, and A. Yalcin, “Nationwide assessment of potential output from wind-powered generators,” \emph{J. Appl. Meteor.}, 15(7), 673--678, 1976.
  \item S. A. Hsu, E. A. Meindl, and D. B. Gilhousen, “Determining the power-law wind-profile exponent under near-neutral stability,” \emph{J. Appl. Meteor.}, 33(6), 757--765, 1994.
  \item Y. Shao and H. Lu, “A simple expression for wind erosion threshold friction velocity,” \emph{JGR: Atmospheres}, 105(D17), 22437--22443, 2000.
  \item J. D. Iversen and B. R. White, “Saltation threshold on Earth, Mars and Venus,” \emph{Sedimentology}, 29(1), 111--119, 1982.
  \item J. F. Kok, E. J. R. Parteli, T. I. Michaels, and D. B. Karam, “The physics of wind-blown sand and dust,” \emph{Rep. Prog. Phys.}, 75(10), 106901, 2012.
  \item I. J. Walker and W. G. Nickling, “Dynamics of secondary airflow and sediment transport over and around barchan dunes,” \emph{JGR: Atmospheres}, 107(D10), 4348, 2002.
  \item G. Sauermann, K. Kroy, and H. J. Herrmann, “A continuum saltation model for sand dunes,” \emph{Phys. Rev. E}, 64(3), 031305, 2001.
  \item H. A. Panofsky and J. A. Dutton, \emph{Atmospheric Turbulence: Models and Methods for Engineering Applications}. Wiley, 1984.
  \item S. B. Pope, \emph{Turbulent Flows}. Cambridge University Press, 2000.
  \item P. Hesp, “Conceptual models of the evolution of transgressive coastal dune field systems,” \emph{Geomorphology}, 199, 138--149, 2013.
  \item W. Pattanapol, “CFD simulation of wind flow over vegetated coastal sand dunes,” PhD thesis, University of Otago, 2010.
\end{enumerate}

\end{document}